\documentclass[%
 aip,pop
 sd,%
 amsmath,amssymb,
 reprint,%
]{revtex4-1}

\usepackage{color}

\usepackage{graphicx}
\usepackage{dcolumn}
\usepackage{bm}

\begin{document}


\title[Perpendicular ion temperature at shock transition region]{
Mach number and plasma beta dependence of the ion temperature
perpendicular to the external magnetic field in the transition region of
perpendicular collisionless shocks
}

\author{Ryo Yamazaki}
\email[Email:]{ryo@phys.aoyama.ac.jp.}
\affiliation{
Department of Physics and Mathematics, 
Aoyama Gakuin University, Sagamihara 252-5258, JAPAN.}

\author{Ayato Shinoda}
\affiliation{
Department of Physics and Mathematics, 
Aoyama Gakuin University, Sagamihara 252-5258, JAPAN.}

\author{Takayuki Umeda}
\email[Email:]{umeda@isee.nagoya-u.ac.jp}
\affiliation{
Institute for Space-Earth Environmental Research, 
Nagoya University, Nagoya 464-8601, JAPAN
}

\author{Shuichi Matsukiyo}
\email[Email:]{matsukiy@esst.kyushu-u.ac.jp}
\affiliation{
Department of Earth System Science and Technology, 
Kyushu University, Kasuga 
816-8580, JAPAN
}



\begin{abstract}

Ion temperature anisotropy is a common feature for (quasi-)perpendicular collisionless shocks.
By using two-dimensional full particle simulations, 
it is shown, that the
ion temperature component perpendicular to the shock magnetic field at the shock foot region
is proportional to the square of the Alfv\'{e}n Mach number divided by the plasma beta.
This result is also explained by a simple analytical argument, in which the reflected ions
get energy from upstream plasma flow.
By comparing our analytic and numerical results, it is also confirmed that the fraction of the reflected ions
hardly depends on the plasma beta and the Alfv\'{e}n Mach number 
when the square of the Alfv\'{e}n Mach number divided by the plasma beta is larger than about 20.

\end{abstract}







\maketitle


%
%




In various kinds of solar-terrestrial, astrophysical and laboratory plasmas, ubiquitous is 
the collisionless shock, at which the upstream kinetic energy of the supersonic plasma flow
dissipates into downstream energy of thermal ions and electrons, waves (turbulence), 
and nonthermal particles.\cite{Balogh_2013,Burgess_2015}
Despite various kinds of studies, detailed processes of the shock dissipation remain to be clarified.
For example, 
we do not fully understand how energies are partitioned between downstream thermal electrons and ions,
although the total pressure of them can be simply predicted by the fluid Rankine-Hugoniot relation.

For supercritical (quasi-)perpendicular shocks, 
a fraction  of incoming ions can be specularly reflected toward the upstream region but gyrates back to the shock 
front.\cite{Woods1971,Sckopke1983,Leroy1983,Gosling1985,Burgess1989,Wikinson1990,Sckopke1990,Lembege_1992}
Such reflected-gyrating ions can gain energy from the motional electric field
of the upstream plasma flow and contribute to the increase of the ion temperature component
perpendicular to the local magnetic field.
Consequently, a large temperature anisotropy arises at the shock foot, exciting waves through the ion temperature
anisotropy instability, which is responsible for the shock ripples.\cite{Winske_1988,Lowe2003}
Electron preheating at the foot also takes place under some conditions.\cite{Woods1971,Hanson2019,Cohen2019}
The ripple further dissipates ions, increasing ion parallel temperature,
and even electron acceleration occurs. \cite{Umeda_2009}
In the downstream region, 
the ion distribution is no longer  non-gyropropic and its structures are smoothed out 
by collisionless gyrophase mixing, resulting in the downstream 
ion heating.\cite{Gedalin1996,Gedalin1997,Gedalin2015a,Gedalin2015b,Ofman2009,Ofman2013}
In order to understand such a multi-step dissipation process across the shock front,
it is important to estimate the initial
ion temperature component perpendicular to the shock magnetic field at the foot region.
In the present study, using the two-dimensional full particle simulation of low-Mach-number, perpendicular, 
rippled and collisionless shocks, we study  the ion perpendicular temperature at the shock foot region.
We show, for the first time, that it is proportional to 
the square of the Alfv\'{e}n Mach number divided by the plasma beta, 
or the square of the sonic Mach number,
which is consistent with the analytical scaling relation.\cite{Sckopke1983}





We perform two-dimensional (2D) simulations of 
perpendicular ($\theta_{B_n} =90^{\circ}$) collisionless shocks 
by using a standard particle-in-cell code.\cite{Umeda_2003}
As in our previous works,\cite{Umeda_2008,Umeda_2006,Umeda_2009}
 the shock is excited by the ``relaxation'' 
between a supersonic and 
a subsonic plasma flows moving in the same direction.
%
%
The initial state consists of the two regions separated by a discontinuity. 
Both regions have spatially uniform distributions of electrons and ions
with different bulk flow velocities, temperatures, and densities,
and they have uniform perpendicular magnetic field with different strength.
The simulation domain is taken in the $x$-$y$ plane 
and an in-plane shock magnetic field ($B_{y0}$) is assumed. 
We apply  a uniform external electric field 
$E_{z0} = u_{x1}B_{y01}/c$~($= u_{x2}B_{y02}/c$) 
in both upstream and downstream regions,  
so that both electrons and ions drift along the $x$ axis. 
Here, $u_{x}$ is the bulk flow velocity, and subscripts ``1'' and ``2'' denote 
``upstream'' and ``downstream'', respectively. 
At the left (right) boundary of the simulation domain in the $x$ direction, 
we inject plasmas with the same quantities 
as those in the initial upstream (downstream) region. 
We use absorbing boundaries 
to suppress non-physical reflection of electromagnetic waves at 
both ends of the simulation domain in the $x$ direction, 
\cite{Umeda_2001} 
while the periodic boundaries are imposed in the $y$ direction.

In the present study, we present results of six simulations runs (A, B, C, D, E and F)
with different upstream conditions.
We summarize in TABLE~I the upstream plasma parameters,  such as
the bulk flow velocity $u_{x1}$, the ratio of the electron plasma frequency to the electron
cyclotron frequency $\omega_{pe1}/\omega_{ce1}$.
For all runs,
we fix  ion-to-electron mass ratio $m_i/m_e=256$, and set $v_{te1}/c=0.1$, where 
$v_{te1}$ and $c$ are the electron thermal velocity upstream and the speed of light, respectively.
In the following, subscripts ``i'' and ``e'' represent 
``ion'' and ``electron'', respectively. 
Then, we obtain the plasma beta
$\beta_1=2(v_{te1}/c)^2(\omega_{pe1}/\omega_{ce1})^2$, 
and $u_{x1}/V_{A1}= (m_i/m_e)^{1/2} (v_{te1}/c)(\omega_{pe1}/\omega_{ce1})(u_{x1}/v_{te1})$, 
where $V_{A1}$ is the upstream Alfv\'{e}n velocity, respectively.
It is assumed that the electrons and ions have the same plasma beta, $\beta_{e1}=\beta_{i1}=\beta_1$, 
and the same isotropic temperature, $T_{e1}=T_{i1}$.
Here, temperatures and thermal velocities are related as $T_{e1} = m_e v_{te1}^2$ and 
$T_{i1} = m_i v_{ti1}^2$.
For given  upstream frequencies $\omega_{pe1}$ and $\omega_{ce1}$,
the initial upstream number density $n_{1} \equiv m_e \omega_{pe1}^2 / 4\pi e^2$
and the initial magnetic field strength $\boldmath{B}_{y01}=m_e \omega_{ce1}/e$  
are derived.
Then, the initial downstream parameters are determined by
solving the shock jump conditions (Rankine-Hugoniot conditions) for 
a magnetized two-fluid isotropic plasma 
consisting of electrons and ions,\cite{Hudson_1970}
assuming $T_{i2}/T_{e2} = 8.0$.

The grid spacing and the time step of the present simulation runs are 
set to be $\Delta x = \Delta y \equiv \Delta = \lambda_{De1}$ 
and $c\Delta t/\Delta = 0.5$, 
where $\lambda_{De}$ is the electron Debye length. 
The total size of the simulation domain is 
$32 l_{i1} \times 6 l_{i1}$, 
where $l_{i1} = c/\omega_{pi1}= (m_i/m_e)^{1/2} (c/v_{te1}) \lambda_{De1}$ is the ion inertial length of the upstream plasma.
We used 25 pairs of electrons and ions per cell in the upstream region 
and 64 pairs of electrons and ions per cell in the downstream region, 
respectively, at the initial state.

\begin{table}[p]
\caption{
\label{tab:result}
Upstream parameters and simulation results.
}
\begin{tabular}{c|lccc|cccc}
\hline\hline
~~Run~~~ &
\multicolumn{4}{c|}{Upstream parameters}
&
\multicolumn{4}{c}{Results}
\\
& ~~$u_{x1}/v_{te1}$~ & ~$\omega_{pe1}/\omega_{ce1}$~~ & ~$\beta_1$~  & ~~$u_{x1}/V_{A1}$~ 
& ~~$M_A$~~ & ~~$M_A{}^2/\beta_1$~~ & $\langle T_{i\perp}^{\rm max}\rangle/T_{i1}$ & ~~$T_{i\perp}^{\rm max}/T_{i1}$ ~~ \\ \hline
A  & ~~1.875     & 2 & 0.08 & 6 & 5.2 & 334  & 183 & 160--205 \\
B  & ~~0.9375   & 4 & 0.32 & 6 & 6.5 & 132  & 63.9 & 53.9--75.4 \\
C  & ~~0.46875 & 8 & 1.28 & 6 & 6.1 & 28.6 & 14.9 & 13.4--17.1 \\
D  & ~~1.25       & 2 & 0.08 & 4 & 4.6 & 265 & 101  & 86.5--123\\
E  & ~~0.625     & 4 & 0.32 & 4 & 4.7 & 70.5 & 30.0 & 26.2--35.5\\
F  & ~~0.3125   & 8 & 1.28 & 4 & 3.9 & 12.5 & 5.01 & 4.80--5.20 \\
\hline\hline
\end{tabular}
\end{table}



\begin{figure}[p]
\centering
\includegraphics[width=8.5cm]{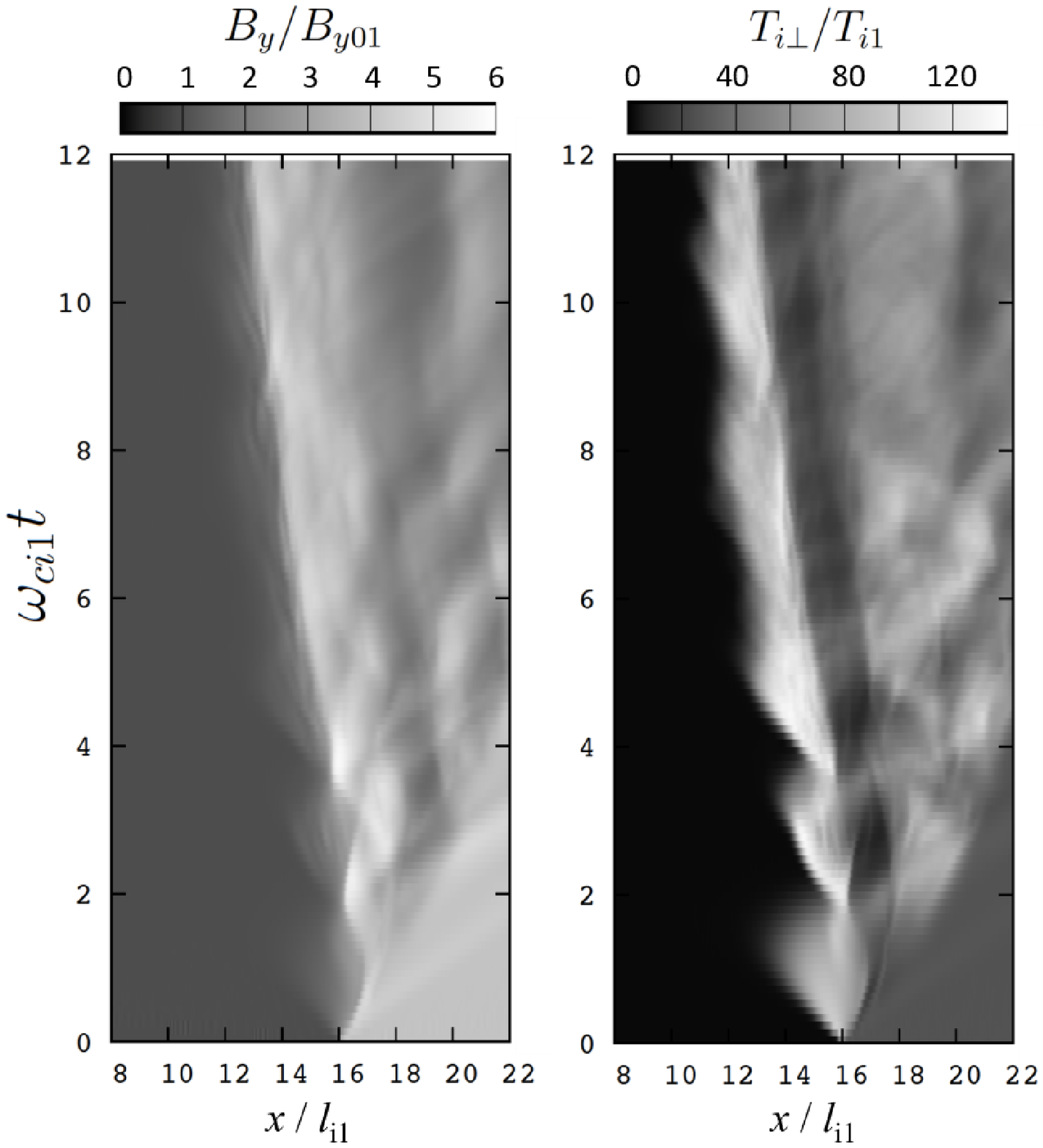}
\vspace{1cm}
\caption{
\label{fig:spatiotemp_Tiperp}
Spatiotemporal diagram of the $y$-component of the magnetic field $B_y$
and ion perpendicular temperature $T_{i\perp}$ for Run~D. Both $B_y$ and $T_{i\perp}$
are averaged over $y$ direction.
Initial unphysical discontinuity is located at $x=16l_{i1}$.
}
\end{figure}

In the following, the ion temperature component perpendicular to the shock magnetic field
is approximated by the arithmetic mean of ion
temperatures in $x$ and $z$ directions, that is, $T_{i\perp}=(T_x+T_z)/2$.
In Fig.~\ref{fig:spatiotemp_Tiperp}, we show spatiotemporal evolution of  $B_y$ (left panel) 
and $T_{i\perp}$ (right panel) for Run~D.
Here both of them are averaged over the $y$ direction.
The initial unphysical disturbance disappears and the growth of the shock ripples ceases 
until $\omega_{ci1}t=7.0$, after which the shock structure seems to be in the quasi-steady state.
Unlike one-dimensional simulation, quasi-periodic reformation seems unclear, although we can still see the
front oscillation.
%
Hence, in the following, we analyze the data  after $\omega_{ci1}t=7.0$ for all runs.

As shown in FIG.~\ref{fig:spatiotemp_Tiperp}, 
the shock front moves leftward in our simulation frame.
Using spatiotemporal diagram of $B_y$, we measure the shock velocity $v_{sh}$,
and obtain the Alfv\'{e}n Mach number, $M_A=(u_{x1}-v_{sh})/V_{A1}$, in the shock-rest frame.
The results for all six runs are shown in TABLE~I.

\begin{figure}[t]
\includegraphics[width=8.5cm]{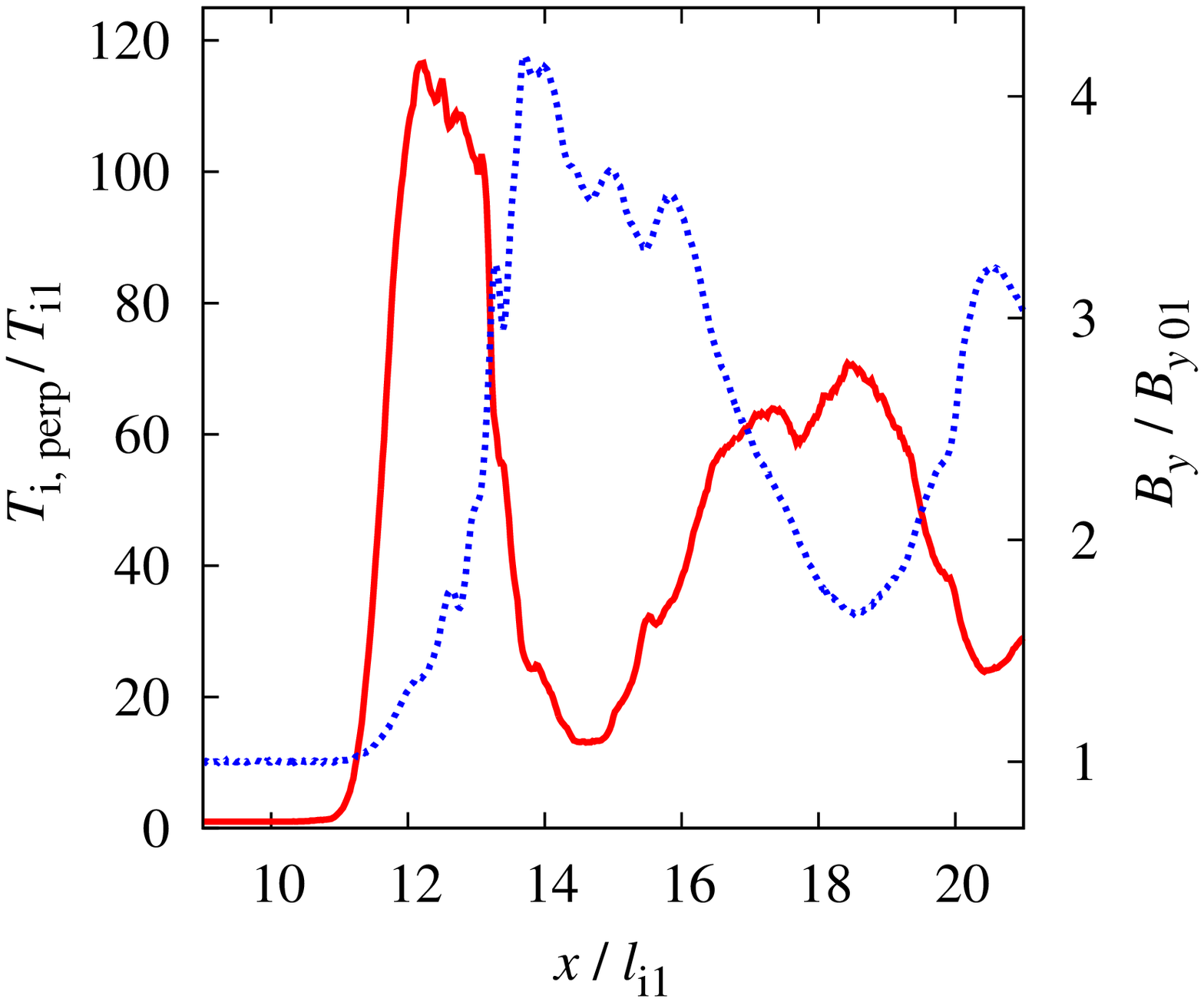}
\vspace{-3.0 cm}
\caption{
\label{fig:Tiperp_By}
Spatial profile of ion perpendicular temperature averaged over $y$ direction, $T_{i\perp}$
(thick solid curve) and the $y$-component of the magnetic field averaged over $y$ direction, $B_y$
(dotted curve) at $\omega_{ci1}t=9.96$ of Run~D.
}
\end{figure}

\begin{figure}
\includegraphics[width=8.5cm]{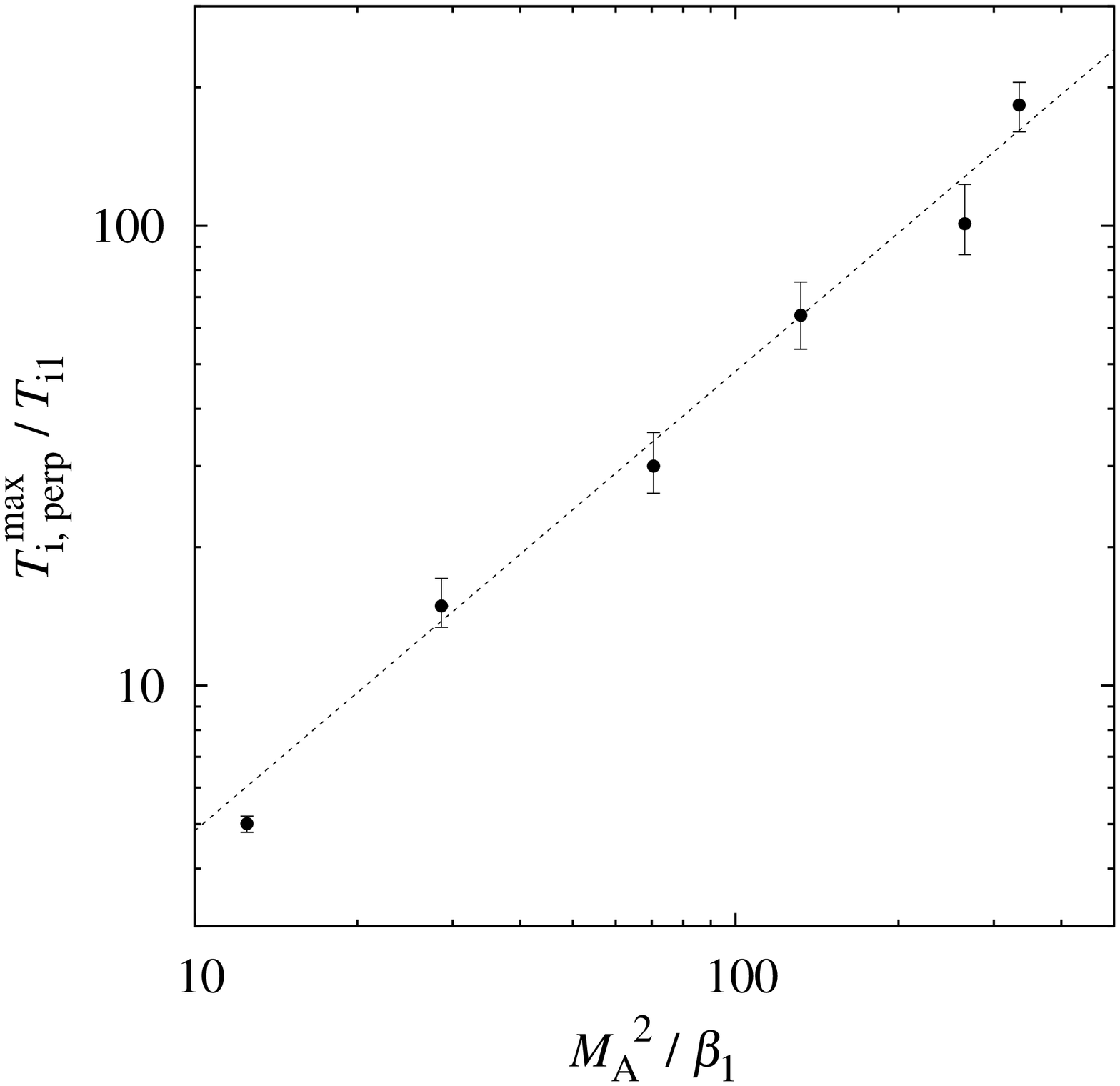}
\vspace{-0.4 cm}
\caption{
\label{fig:Tiperp_Mach}
Ion perpendicular temperature $T_{i\perp}^{\rm max}$ as a function of $M_{A}{}^2/\beta_1$.
Triangles represent $\langle T_{i\perp}^{\rm max}\rangle/T_{i1}$, while
error bars indicate the maximum and minimum values of $T_{i\perp}^{\rm max}/T_{i1}$,
which are given in TABLE~\ref{tab:result}. All the values are obtained for
$7<\omega_{ci1}t<12$.
Dotted line represents best-fitted linear relation, $T_{i\perp}^{\rm max}/T_{i1}=0.48 \times M_{A}{}^2/\beta_1$.
}
\end{figure}

We extract the representative value of $T_{i\perp}$ in the transition layer for each run 
as follows.
 First, we make a snapshot of $T_{i\perp}$ which is averaged over $y$ direction,
 and find its maximum value, $T_{i\perp}^{\rm max}$, for each time step.
As shown in FIG.~\ref{fig:Tiperp_By},
one can find that $T_{i\perp}$ has a maximum at the foot region.
This fact is also true  in arbitrary epoch.
%
%
%
The value of $T_{i\perp}^{\rm max}$ changes with time.
Then, the time mean of $T_{i\perp}^{\rm max}$  for $7\le\omega_{ci1}t\le12$
is obtained.
For example, we obtain the average value $T_{i\perp}^{\rm max}/T_{i1}=101$ and
the maximum and minimum values are 123 and 86.5, respectively.
In the same way, the average, maximum, and minimum values of $T_{i\perp}^{\rm max}$ are 
obtained for other runs.
The results are summarized in TABLE~I.
In FIG.~\ref{fig:Tiperp_Mach}, $T_{i\perp}^{\rm max}/T_{i1}$ is shown as a function of $M_{A}{}^2/\beta_1$.
The simulation results seem to lie on the line, $T_{i\perp}^{\rm max}/T_{i1}\approx0.5 \times M_{A}{}^2/\beta_1$.

In the following,
we derive a simple analytical formula to have the ion perpendicular temperature at the shock foot.
Although similar formulae have been already derived, \cite{Sckopke1983,Gosling1985,Burgess1989}
our final equation, Eq.~(\ref{eq:Tperp2}), is 
in an excellent agreement with 
our simulation results.
In our simulation frame,  shock front moves at a velocity $v_{sh}$, and
upstream and downstream bulk velocities are typically $u_{x1}$ and $u_{x2}$, respectively.
The incoming ions are adiabatically heated at the shock foot.
They have the perpendicular temperature, $[C^{\gamma-1}/(\gamma-1)]m_iv_{ti1}^2$, where
$C\sim(n_{i,f}/n_{i1})$
is a compression factor, and $\gamma$ and $n_{i,f}$ are the adiabatic index ($\gamma=2$ in our case) and 
 the typical value of the density at the foot, respectively.
Since they are also decelerated and due to the mass flux conservation,
their bulk velocity becomes $u_{{\rm (in)}}\approx v'_{sh}/C$ measured in the rest frame of the shock front, 
where  $v_{sh}' =u_{x1}-v_{sh}=M_AV_{A1}$.
Next, a part of them are reflected.
The bulk velocity of the reflected ions is $u_{{\rm (ref)}}\approx -v'_{sh}/C$ in the shock rest frame. \cite{Leroy1983}
%
Hence, we have the velocity difference between the incoming and the reflected ions,
\begin{eqnarray}
\Delta u\approx u_{{\rm (in)}}-u_{{\rm (ref)}}\approx \frac{2v'_{sh}}{C}~~,
\label{eq:veldiff}
\end{eqnarray}
at the shock foot.
A large fraction of  energy (per ion), $(m_i/2)|\Delta u|^2$, is consumed for increasing the ion
perpendicular temperature.

Here, we consider a simple analytical model to estimate $T_{i\perp}$ at the foot region.
The ion distribution function there is written as
\begin{eqnarray}
f_{\rm tot}(v_x)=f_{\rm (in)}(v_x)+f_{\rm (ref)}(v_x)~~,
\end{eqnarray}
where the first and the second terms in r.h.s. describe the incoming and reflected components,
respectively, and they have the number density $N_{\rm (k)}$,
bulk velocity $u_{\rm (k)}$, and temperature $T_{\rm (k)}$ as
\begin{eqnarray}
N_{\rm (k)} &=& \int f_{\rm (k)}dv_x ~~,\\
u_{\rm (k)} &=& \frac{1}{N_{\rm (k)}}\int v_x f_{\rm (k)} dv_x ~~,\\
T_{\rm (k)} &=& \frac{m_i}{N_{\rm (k)}}\int(v_x-u_{\rm (k)})^2f_{\rm (k)}dv_x~~,
\end{eqnarray}
respectively. A subscript ${\rm (k)}={\rm (in)}$, (ref)  denotes each component.
Then, it is natural to approximate $T_{i\perp}$ as
\begin{eqnarray}
T_{i\perp}\approx T_x = \frac{m_i}{N_{\rm tot}}\int (v_x-\bar{u})^2 f_{\rm tot} dv_x~~,
\end{eqnarray}
where $N_{\rm tot}$ and $\bar{u}$ are given by
\begin{eqnarray}
N_{\rm tot}&=&\int f_{\rm tot}dv_x=N_{\rm (in)}+N_{\rm (ref)}~~,\\
\bar{u} &=&\frac{1}{N_{\rm tot}}\int v_x f_{\rm tot}dv_x~~,
\end{eqnarray}
respectively.
When we introduce the fraction of the reflected ions as
$r=N_{\rm (ref)}/N_{\rm tot}$, then we get
\begin{eqnarray}
T_{i\perp} &=& T_{\rm (in)} + r(T_{\rm (ref)}-T_{\rm (in)}) \nonumber\\
&& \ \ +m_i[(1-r)u_{\rm (in)}^2+ru_{\rm (ref)}^2-\bar{u}^2]~~,
\label{eq:Tiperp1}
\end{eqnarray}
and $\bar{u}=(1-r)u_{\rm (in)}+ru_{\rm (ref)}$.
Assuming $T_{\rm (in)}=T_{\rm (ref)}$ as in one of our previous simulation study\cite{Umeda_2012b}
and eliminating $\bar{u}$, we can rewrite Eq.~(\ref{eq:Tiperp1}) as
\begin{eqnarray}
T_{i\perp}=T_{\rm (in)}+r(1-r)m_i(u_{\rm (in)}-u_{\rm (ref)})^2~~.
\end{eqnarray}
Using Eq.~(\ref{eq:veldiff}) together with $T_{\rm (in)}=CT_{i1}$ and $m_iv'_{sh}{}^2/T_{i1}=2M_A{}^2/\beta_1$,
we finally obtain
\begin{eqnarray}
\frac{T_{i\perp}}{T_{i1}}= C+\frac{8r(1-r)}{C^2}\frac{M_A{}^2}{\beta_1}~~.
\label{eq:Tperp2}
\end{eqnarray}
The first term in r.h.s of Eq.~(\ref{eq:Tperp2}) is important for the case of low $M_{A}{}^2/\beta_1$ only.
The compression factor $C$ is slightly larger than unity for such shocks.
For large $M_{A}{}^2/\beta_1$, the second term dominates r.h.s of Eq.~(\ref{eq:Tperp2}),
so that we can explain our numerical result,
$T_{i\perp}^{\rm max}/T_{i1} \propto M_{A}{}^2/\beta_1$, shown in FIG.~\ref{fig:Tiperp_Mach},
if the factor $8r(1-r)/C^2$ in Eq.~(\ref{eq:Tperp2}) hardly depends on $M_{A}{}^2/\beta_1$.
%
%
The fraction of the reflected ions is typically $r \approx 0.3$ and
varies from 0.2 to 0.4 during non-stationary processes at the shock front.
On the other hand, the value of $C$ is less variable and ranges between 1.0 and 1.1.
Then, one can see $8r(1-r)/C^2=1.1-1.9$.
This is a factor of a few larger than estimated from FIG.~\ref{fig:Tiperp_Mach}.
Indeed, our analytical formula, Eq.~(\ref{eq:Tperp2}), gives the upper bound of $T_{i\perp}$,
because the free energy $m_i(\Delta u)^2$ goes not only to $T_{i\perp}$ but also to the
thermal energy of reflected ions and waves excited in the shock transition layer.\cite{Umeda_2014}
In practice, one can see that $T_{i\perp}$ becomes larger if $T_{\rm (ref)}>T_{\rm (in)}$
[see Eq.~(\ref{eq:Tiperp1})].
Note that the  fraction of the reflected ions, $r$, estimated in the previous analytical works \cite{Leroy1983,Wikinson1990,Hada2003} 
is slightly smaller than that obtained from our simulations. 

In the present study, we focus on $T_{i\perp}$ only.
On the other hand, the parallel component of ion temperature ($T_{i\parallel}\approx T_y$)
is much smaller at the foot region.
Therefore, the total ion temperature,
$T_i =(T_x+T_y+T_z)/3 =(2T_{i\perp}+T_{i\parallel})/3$,
is approximated as $T_i\approx (2/3)T_{i\perp}$.




Using two-dimensional full particle simulations,
we have shown that the ion perpendicular temperature at the foot of the supercritical perpendicular
collisionless shocks 
is proportional to $M_{A}{}^2/\beta_1$, or the square of the sonic Mach number.
This fact will give us a simple estimate of the energy partition between downstream thermal ions and electrons, 
although further study is necessary.
The ion heating at the foot region of (quasi-)perpendicular shocks has been extensively investigated
by many authors, using
mainly spacecraft observations,\cite{Sckopke1983,Sckopke1990}
(semi-)analytical 
studies,\cite{Leroy1983,Sckopke1983}
and one-dimensional hybrid simulations.\cite{Burgess1989}
In this paper, we have extended such studies by using two-dimensional full
particle simulations which can better capture various kinetic effects including
wave excitations and plasma heating in the direction tangential to the shock front. 
More specifically, we have demonstrated in this paper that our analytical scaling relation 
Eq.~(\ref{eq:Tperp2})  is in excellent agreement with 
two-dimensional full particle simulations of rippled shocks.
This result also indicates that the dependence of fraction of ion reflection 
on the plasma beta and the Alfv'{e}n Mach number is small 
when $M_A^2/\beta$ is larger than about 20,  
which is consistent with our simulation results.


\begin{acknowledgments}


The authors would like to thank Yutaka Ohira for helpful comments.
%
%
The computer simulations were performed on the CIDAS supercomputer system 
at the Institute for Space-Earth Environmental Research in Nagoya University 
under the joint research program. 
This work was partly supported by JSPS KAKENHI Grants: 15K05088, 18H01232 (RY), 2628704, 19H01868 (TU).

\end{acknowledgments}




\begin{thebibliography}{29}



\bibitem{Balogh_2013}
A. Balogh and R. A. Treumann,
Physics of Collisionless Shocks
(Springer, New York, 2013)

\bibitem{Burgess_2015}
D. Burgess and M. Scholer,
Collisionless Shocks in Space Plasmas
(Cambridge University Press, Cambeidge, 2015)


\bibitem{Woods1971}
L. C. Woods,
Plasma Phys. \textbf{13}, 289 (1971).

\bibitem{Leroy1983}
M. M. Leroy,
Phys. Fluids. \textbf{26}, 2742 (1983).

\bibitem{Sckopke1983}
N. Sckopke, G. Paschmann, S. J. Bame, and J. T. Gosling,
J. Geophys. Res. \textbf{88}, 6121 (1983).

\bibitem{Gosling1985}
J. T. Gosling and A. E. Robson,
In collisionless shocks in the heliosphere: Reviews of surrent research,  \textbf{35}, 141.
Washington, DC: American Geophysical Union  (1985).

\bibitem{Burgess1989}
D. Burgess, W. P. Wilkinson, and S. J. Schwartz,
J. Geophys. Res. \textbf{94}, 8783 (1989).

\bibitem{Wikinson1990}
W. P. Wilkinson and S. J. Schwartz,
Planet Space Sci. \textbf{38}, 419 (1990).

\bibitem{Sckopke1990}
N. Sckopke, G. Paschmann, A. L. Brinca, C. W. Carlson, and H. L\"{u}hr, 
J. Geophys. Res. \textbf{95}, 6337 (1990).

\bibitem{Lembege_1992}
B. Lembege and P. Savoini, 
Phys. Fluids B \textbf{4}, 3533
(1992). 


\bibitem{Winske_1988}
D. Winske and K. B. Quest, 
J. Geophys. Res. \textbf{93}, 9681 
(1988).

\bibitem{Lowe2003}
R. E. Lowe and D. Burgess,
Ann. Geophys. \textbf{21}, 671 (2003).

\bibitem{Hanson2019}
E. L. M. Hanson, O. V. Agapitov, F. S. Mozer, V. Krasnoselskikh,
S. D. Bale, L. Avanov, Y. Khotyaintsev, and  B. Giles
Geophys. Res. Lett. \textbf{46}, 2381 (2019).

\bibitem{Cohen2019}
I. J. Cohen, S. J. Schwartz, K. A. Goodrich, N. Ahmadi, R. E. Ergun, 
S. A. Fuselier, M. I. Desai, E. R. Christian, D. J. McComas, G. P. Zank, 
et al., 
J. Geophys. Res. \textbf{124}, 3961 (2019).

\bibitem{Umeda_2009}
T. Umeda, M. Yamao, and R. Yamazaki, 
Astrophys. J. \textbf{695}, 574 
(2009). 


\bibitem{Gedalin1996}
M. Gedalin,
J. Geophys. Res. \textbf{101}, 15569 (1996).

\bibitem{Gedalin1997}
M. Gedalin,
Geophys. Res. Lett. \textbf{24}, 2511 (1997).

\bibitem{Gedalin2015a}
M. Gedalin,
J. Plasma Phys. \textbf{81}, 905810603 (2015).

\bibitem{Gedalin2015b}
M. Gedalin,
Phys. Plasmas. \textbf{22}, 072301 (2015).




\bibitem{Ofman2009}
L. Ofman, M. Balikhin, C. T. Russell, and M. Gedalin,
J. Geophys. Res. \textbf{114}, A09106 (2009).


\bibitem{Ofman2013}
L. Ofman and M. Gedalin,
J. Geophys. Res. \textbf{118}, 1828 (2013).




\bibitem{Umeda_2003}
T. Umeda, Y. Omura, T. Tominaga, and H. Matsumoto, 
Comput. Phys. Commun. \textbf{156}, 73 
(2003). 

\bibitem{Umeda_2006}
T. Umeda and R. Yamazaki, 
Particle simulation of a perpendicular collisionless shock: 
A shock-rest-frame model, 
Earth Planets Space \textbf{58}, e41 
(2006). 


\bibitem{Umeda_2008}
T. Umeda, M. Yamao, and R. Yamazaki, 
Astrophys. J. \textbf{681}, L85 
(2008). 

\bibitem{Umeda_2001}
T. Umeda, Y. Omura, and H. Matsumoto, 
Comput. Phys. Commun. \textbf{137}, 286 
(2001). 

\bibitem{Hudson_1970}
P. D. Hudson, 
Planet. Space Sci. \textbf{18}, 1611 
(1970). 

\bibitem{Umeda_2012b}
T. Umeda, Y. Kidani, S. Matsukiyo, and R. Yamazaki, 
Phys. Plasmas \textbf{19}, 042109 
(2012). 

\bibitem{Umeda_2014}
T. Umeda, Y. Kidani, S. Matsukiyo, and R. Yamazaki, 
Phys. Plasmas \textbf{21}, 022102 
(2014). 


\bibitem{Hada2003}
T. Hada, M. Oonishi, B. Lemb\`{e}ge, and P. Savoini, 
J. Geophys. Res. \textbf{108}, 1233 (2003).



\end{thebibliography}



\end{document}